\documentclass{article}
\usepackage{spconf,amsmath,graphicx}
\usepackage{booktabs}
\usepackage{multirow}
\usepackage{float}
\usepackage[font=small,skip=4pt]{caption}

\title{Improving Speech Emotion Recognition \\with Unsupervised Speaking Style Transfer}
\name{Leyuan Qu$^{1}$, Wei Wang$^{2}$, Cornelius Weber$^{3}$, Pengcheng Yue$^{1}$, Taihao Li$^{1}$*, Stefan Wermter$^{3}$ \thanks{*Corresponding author (email: lith@zhejianglab.com)}}
\address{$^{1}$Zhejiang Lab, China, $^{2}$Xinjiang University, China, $^{3}$University of Hamburg, Germany}
\begin{document}
%
\maketitle
%

%

Humans can effortlessly modify various prosodic attributes, such as the placement of stress and the intensity of sentiment, to convey a specific emotion while maintaining consistent linguistic content. Motivated by this capability, we propose EmoAug, a novel style transfer model designed to enhance emotional expression and tackle the data scarcity issue in speech emotion recognition tasks. EmoAug consists of a semantic encoder and a paralinguistic encoder that represent verbal and non-verbal information respectively. Additionally, a decoder reconstructs speech signals by conditioning on the aforementioned two information flows in an unsupervised fashion. Once training is completed, EmoAug enriches expressions of emotional speech with different prosodic attributes, such as stress, rhythm and intensity, by feeding different styles into the paralinguistic encoder. EmoAug enables us to generate similar numbers of samples for each class to tackle the data imbalance issue as well. Experimental results on the IEMOCAP dataset demonstrate that EmoAug can successfully transfer different speaking styles while retaining the speaker identity and semantic content. Furthermore, we train a SER model with data augmented by EmoAug and show that the augmented model not only surpasses the state-of-the-art supervised and self-supervised methods but also overcomes overfitting problems caused by data imbalance. Some audio samples can be found on our demo website\footnote{https://leyuanqu.github.io/EmoAug/}.

%
\section{Introduction}
\label{sec:intro}
Speech Emotion Recognition (SER) aims at recognizing and understanding human emotions from spoken language, which can significantly benefit and promote the experience of human-machine interaction.
Plenty of studies have investigated SER in recent years~\cite{wagner2023dawn}.
However, the performance of SER is constrained by the lack of large-scale labelled datasets. Most speech emotion datasets are collected under simulated or elicited scenarios, since capturing natural and spontaneous emotional speech is very challenging. Furthermore, different people perceive emotions differently, which may result in ambiguity in data labeling, especially in emotions with weak intensity~\cite{abbaschian2021deep}. Moreover, data imbalance of different emotions is another problem that can lead to model overfitting to some frequent emotions, such as ``Neutral".



Data augmentation is treated as a promising method to address these issues. Most of the augmentation methods used in SER are borrowed from Automatic Speech Recognition (ASR), such as SpecAugment~\cite{park2019specaugment}, Vocal Tract Length Perturbation (VTLP)~\cite{jaitly2013vocal}, speed perturbation~\cite{ko2015audio}, pitch shift or noise injection. However, while the variations in pitch or speed do not change the semantic content, they may have an effect on emotion expressions. For instance, sad emotions are often conveyed at slow speed while angry emotions tend to be expressed fast. 

Alternatively, different variants of models are adopted to generate intermediate emotional features or alter carried emotions while keeping speech content unchanged, for instance, star Generative Adversarial Networks (GANs)~\cite{rizos2020stargan}, CycleGAN~\cite{bao2019cyclegan} and global style token~\cite{wang2018style}. However, the generated intermediate features are not easy to evaluate intuitively. In addition, prosody expression is strongly associated with speech content, and the altered emotions by GANs may cause conflicts or ambiguity between speech prosody and content.


Humans can easily alter different prosody attributes, such as stress position and sentiment intensity, to express a given emotion with invariant linguistic content~\cite{schuller2013computational}. For example, when expressing sad emotions with the semantic content ``I am not happy today", one can emphasize ``not happy" or put stress on ``today". Inspired by this capability, we propose EmoAug to vary prosody attributes and augment emotional speech while keeping emotions unchanged. EmoAug is trained in an unsupervised manner, which requires neither paired speech nor emotion labels. The main contributions of this paper are as follows: 

\begin{figure*}[th]
\setlength{\belowcaptionskip}{-0.5cm}
  \centering
  \includegraphics[width=15cm]{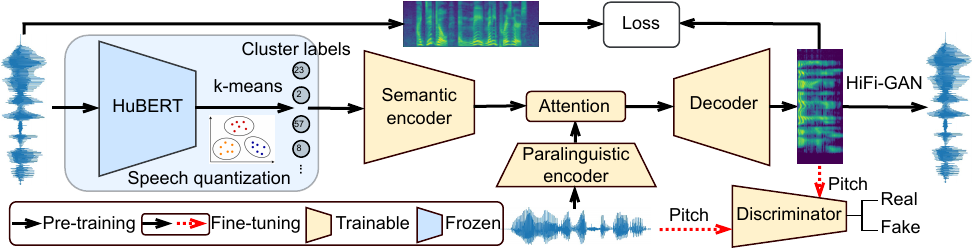}
  \caption{Overview of EmoAug. The model comprises two encoders and an attention-based decoder. In the pre-training phase, the decoder reconstructs input mel-spectrograms using representations acquired from the semantic and paralinguistic encoders. The loss function is computed by measuring the Mean Square Error (MSE) between the generated and original mel-spectrograms. During fine-tuning, style transfer is performed by directly substituting the input of the paralinguistic encoder with a target speaking style reference. Additionally, to enhance the quality of the converted audio, a discriminator is employed to differentiate between real and generated pitch contours.}
  \label{fig:model}
\end{figure*}

\vspace{-0.3cm}
\begin{enumerate}
\item We propose a novel unsupervised speaking style transfer model which enriches emotion expressions by altering stress, rhythm and intensity while keeping emotions, semantics and speaker identity invariant.
\vspace{-0.3cm}
\item We implement quantized units to represent semantic speech content instead of using a well-trained ASR model, which enables us to work on emotional audio without text transcriptions.
\vspace{-0.3cm}
\item SER models trained with EmoAug outperform the state-of-the-art models by a large margin, effectively overcoming overfitting issues caused by data scarcity and class imbalance.
\end{enumerate}




\vspace{-0.6cm}

\section{Method}

An overview of EmoAug is shown in Fig.~\ref{fig:model}, which consists of a speech quantization module, a semantic encoder, a paralinguistic encoder and an attention-based decoder. An additional discriminator is utilized during the fine-tuning phase.



\vspace{-0.3cm}
\subsection{Speech Quantization and Semantic Encoder}
Acquiring semantic speech information typically relies on a proficient ASR system. Nevertheless, training a reliable ASR model proves challenging due to the extensive labeled data required. In addition, ASR models are not robust to unseen noise or emotional speech~\cite{lakomkin2018robustness}. Therefore, we employ HuBERT representations~\cite{hsu2021hubert} to capture semantic speech content. These representations are acquired through self-supervised training, eliminating the need for human annotations in ASR training. HuBERT empowers us to address emotional speech and diverse styles that might substantially compromise the performance of an ASR model.
As shown in Eq.\ref{eq1}, the input speech signal $x = (x_0,...,x_t)$ is firstly embedded into continuous vectors by the pre-trained HuBERT\footnote{https://huggingface.co/facebook/hubert-base-ls960}, followed by the k-means algorithm that quantizes the continuous speech representations into discrete cluster labels $u=(u_0,...,u_t)$, e.g. ``23, 23, 2, 2, ..., 57".

\vspace{-0.3cm}
\begin{equation}
\setlength{\abovedisplayskip}{0.5pt}
\setlength{\belowdisplayskip}{0.5pt}
u=k\mbox{-}means(HuBERT(x))
\label{eq1}
\end{equation}
\vspace{-0.4cm}

We investigated how different vocabulary sizes of HuBERT impact in model performance.
In this study, we employed a vocabulary size of 200 clusters, which yielded considerably improved performance compared to using 50 or 100 classes.
To refine the semantic content, we proceed to eliminate repetitions and filter out tempo information ($u \rightarrow \widetilde{u}$), e.g. ``23, 23, 2, 2, ..., 57" $\rightarrow$ ``23, 2, ..., 57". After quantization, we map the cluster labels into latent representations with the semantic encoder (\textit{Sem}) which is comprised of three 512-channel Conv1D layers with kernel width of 5 and padding size of 2, and one bidirectional Long Short-Term Memory (LSTM) with 256 dimensions to capture local and global context information, respectively.


\vspace{-0.3cm}
\subsection{Paralinguistic Encoder}

The paralinguistic encoder (\textit{Par}) aims to learn utterance-level non-verbal information from input audio $x$, which contains speaking styles, emotion states, speaker identities, and so on. It is based on the ECAPA-TDNN model~\cite{DesplanquesTD20} which has been proposed for the task of speaker verification.
The model begins with one Conv1D layer, a ReLU function, and Batch Normalization (BN), followed by three SE-Res2 blocks. Residual connections between the SE-Res2 blocks deliver different level outputs to the feature aggregation layer (Conv1D+ReLU). Subsequently, the aggregated outputs are dynamically weighed by the attentive statistics pooling layer (Conv1D+Tanh+Conv1D+Softmax), and then mapped to a fixed dimension by the last Fully Connected (FC) layer. We initialize ECAPA-TDNN with the pre-trained model\footnote{https://huggingface.co/speechbrain/spkrec-ecapa-voxceleb}. The attentive statistics pooling, fixed dimension mapping and weight initialization prevent semantic information to leak from the paralinguistic encoder.

\vspace{-0.3cm}
\subsection{Decoder}

Our decoder is based on the Tacotron2~\cite{shen2018natural} which uses location-aware attention (\textit{Att})~\cite{chorowski2015attention} to connect the encoders and the decoder (see Eq.~\ref{eq2}).

\vspace{-0.3cm}
\begin{equation}
\setlength{\abovedisplayskip}{0.5pt}
\setlength{\belowdisplayskip}{0.5pt}
\widetilde{m}=Dec(Att(Sem(\widetilde{u}),Par(x)))
\label{eq2}
\end{equation}
\vspace{-0.4cm}

The decoder (\textit{Dec}) generates one frame per time step in an auto-regressive fashion. Two FC layers map the ground-truth mel-spectrograms $x$ into latent representations that are used by an LSTM module for teacher-forcing training. Finally, a FC layer maps the intermediate features to the dimension of input mel-spectrograms. The generated mel-spectrograms $\widetilde{m}$ are then transformed to waveforms $\widetilde{x}$ by HiFiGAN~\cite{kong2020hifi}.




\vspace{-0.3cm}
\begin{equation}
\setlength{\abovedisplayskip}{0.1cm}
\setlength{\belowdisplayskip}{0.1cm}
\widetilde{x}=HiFiGAN(\widetilde{m})
\label{eq3}
\end{equation}
\vspace{-0.8cm}

\subsection{Discriminator}

After pre-training, we observed when feeding a different reference audio to the paralinguistic encoder for style transfer, the generated speech signals exhibit some distortions. To enhance the quality of the generated speech, we implement a discriminator to differentiate between the genuine and synthesized speech. Importantly, the discriminator exclusively differentiates pitch changes while preserving semantic content.
The discriminator starts with three convolutional blocks (Conv1D+ReLU+BN+dropout), followed by two linear projection layers. 




\vspace{-0.3cm}
\subsection{Speaking Style Transfer}
\label{style-transfer}

We denote $X^{s,e}$ as all utterances with emotion $e$ uttered by speaker $s$. After pre-training, as shown in Eq.~\ref{eqtrans1}, speaking style transfer can be achieved by directly replacing the paralinguistic encoder input $x^{s,e}$ with a target speaking style $y^{s,e}$, where $x^{s,e}\in X^{s,e}$, $y^{s,e}\in Y^{s,e}$ and $Y^{s,e}=X^{s,e}-\{x^{s,e}\}$.

\vspace{-0.3cm}
\begin{equation}
\setlength{\abovedisplayskip}{0.5pt}
\setlength{\belowdisplayskip}{0.5pt}
\widetilde{m}^{s,e}=Dec(Att(Sem(\widetilde{u}),Par(y^{s,e})))
\label{eqtrans1}
\end{equation}
\vspace{-0.3cm}

Consequently, the converted mel-spectrograms $\widetilde{m}^{s,e}$ retain the same speaker identity, semantic content and emotions as the original audio $x^{s,e}$, but deliver different rhythms or intensities transferred from $y^{s,e}$. In addition, we generate different numbers of samples for each emotion to counter data imbalance.

\vspace{-0.3cm}
\section{Experimental Setups}
\vspace{-0.2cm}
\subsection{Datasets}

We utilize the LRS3-TED and IEMOCAP datasets for pre-training and fine-tuning respectively.

\vspace{-0.2cm}
\begin{itemize}
\item \textbf{LRS3-TED}~\cite{afouras2018lrs3} is comprised of over 400 hours of video by more than 5000 speakers from TED and TEDx talks with spontaneous speech. LRS3-TED is collected with various speaking styles and emotions in a variety of acoustic scenes, which will help the model learn rich paralinguistic changes. 
\vspace{-0.2cm}


\item \textbf{IEMOCAP}~\cite{busso2008iemocap} is a multimodal emotion dataset recorded by 10 actors in a fictitious scenario. We follow the settings used in previous work~\cite{guo2021representation, tarantino2019self, zou2022speech}, in which leave-one-session-out is adopted with 5-fold cross-validation. One session is left for testing and another one is used for validation, while the rest of the three sessions is utilized for training in each round. Four types of emotions (\textit{happy, sad, angry} and \textit{neutral}) with a total of 5531 utterances are considered.
\end{itemize}
\vspace{-0.2cm}



\vspace{-0.3cm}
\subsection{Evaluation Metrics}

We report the experimental results with Weighted Accuracy (WA) and Unweighted Accuracy (UA). 

\vspace{-0.3cm}
\subsection{EmoAug Training}

We first pre-train EmoAug on LRS3-TED 
and then fine-tune it on IEMOCAP for style transfer. 
During pre-training, the Adam optimizer is used with a weight decay rate of 1e-6. The initial learning rate is 1e-3, which is decayed with a factor of 0.9 after every 5000 iterations. In addition, gradient clipping with a threshold of 1.0, early stopping and scheduled sampling are adopted to avoid overfitting. 




After pre-training, we fine-tune the model with a discriminator on the IEMOCAP dataset with small learning rates of 1e-5 and 1e-4 to improve the quality of generated speech. 



\vspace{-0.3cm}
\subsection{SER Model Training}

We perform SER by adding one additional FC layer on top of the HuBERT model which is pre-trained on large-scale unlabelled data by self-supervised learning. During training, different learning rates are used in the HuBERT model (1e-5) and the FC layer (1e-4) to retain the low-layer representations and enable the last FC layer to fit to the specific dataset.

\vspace{-0.3cm}
\section{Experimental Results and Discussion}

\subsection{Comparison of Different Augmentation Times}

We augment each utterance $N$ times by transferring speaking styles from $N$ randomly selected utterances that belong to the same speaker with the same type of emotion. We report the effect of different augmentation times and the discriminator on SER utilizing representations from HuBERT. 

\begin{figure}[H]
  \centering
  \setlength{\abovecaptionskip}{0.cm}
  \includegraphics[width=\linewidth]{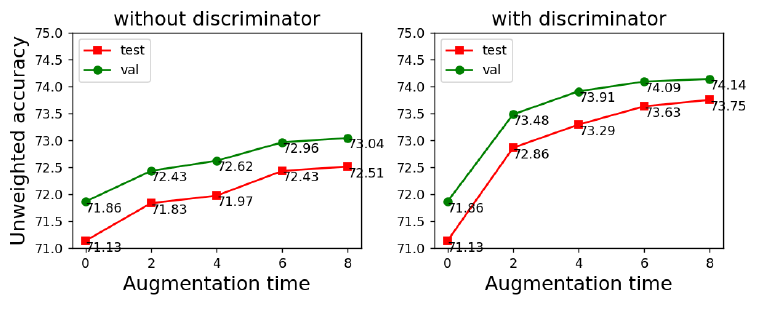}
  \caption{The effect of different augmentation times on SER when training EmoAug with or without the discriminator.}
  \label{fig:aug-times}
  \setlength{\belowcaptionskip}{0.cm}
\vspace{-0.1cm}
\end{figure}

As can be seen in Fig.~\ref{fig:aug-times}, considerable increases occur on UA along with the increase of augmentation time, where $0$ means training SER models with only raw audio. Furthermore, the discriminator significantly enhances the quality of the generated speech, resulting in a substantial improvement in emotion recognition. The figure reveals that the speaking styles transferred by EmoAug effectively enrich the emotion expression on prosody and greatly enhance SER performance.

\vspace{-0.3cm}

\vspace{-0.3cm}
\begin{figure*}[th]
  \centering
  \includegraphics[width=16cm]{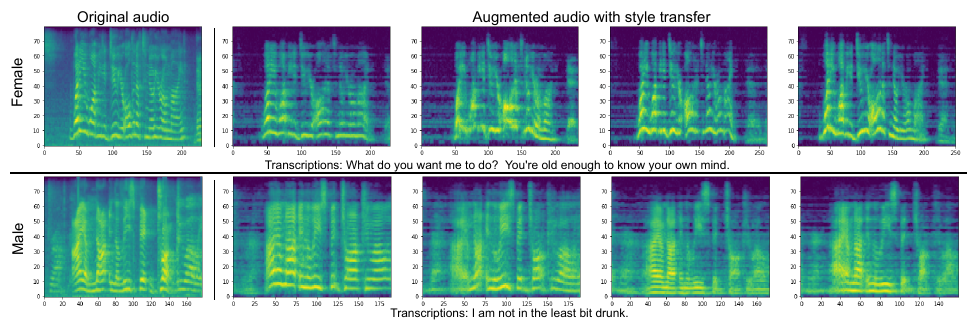}
  \caption{A comparison of original and augmented mel-spectrograms with the \textit{angry} emotion where the styles are transferred from the same speaker with the same emotion as the original audio. Different stress patterns, rhythms and intonations are shown in the augmented audio.}
  \label{fig:mel}
 \vspace{-0.5cm}
\end{figure*}

In addition, we also visualize the original and generated mel-spectrograms.
As shown in Fig.~\ref{fig:mel}, in comparison to the original audio, EmoAug successfully transfers different speaking styles to the source audio while keeping the semantic information invariant. By listening to the generated audio, we found that EmoAug can effectively augment the expression of source emotion by varying stress positions, intonations or even the intensity of emotions. 
We recommend readers to listen to the audio samples on our demo website.



\vspace{-0.3cm}
\subsection{Model Performance with and without Augmentation}

\begin{figure}[H]
  \centering
  \setlength{\abovecaptionskip}{0.cm}
  \includegraphics[width=8cm]{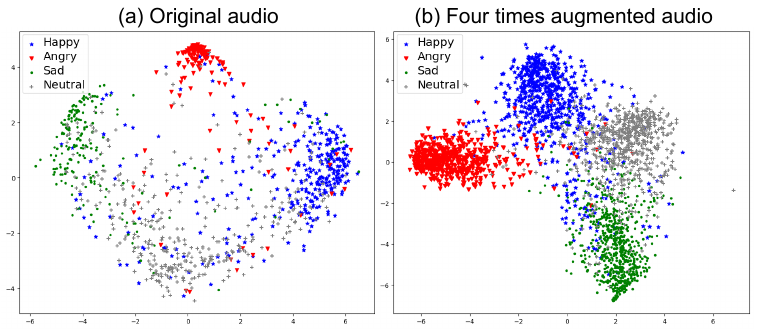}
  \caption{Comparison of models trained with original data (a) and augmented data (b), visualized with t-SNE.}
  \label{fig:matrix}
\end{figure}

We select the utterances from session one (two speakers) of IEMOCAP and visualize the embeddings from the penultimate layer of the model trained with only original data and with four times augmented data. As depicted in Fig.~\ref{fig:matrix} (a), the model trained only with original audio struggles to clearly separate the representations of neutral, sad and happy emotions. However, as shown in Fig.~\ref{fig:matrix} (b), EmoAug successfully transfers speaking styles and enriches sample diversity, resulting in more precise and distinguishable model representations when trained with augmented audio.

\vspace{-0.2cm}
\vspace{-0.2cm}

\vspace{-0.6cm}
\subsection{Comparison with Previous Work}

We compare our methods with previous supervised and self-supervised models in Table~\ref{table:comparison-with-other}. We also reimplement and augment IEMOCAP with emotional voice conversion models, CycleGAN~\cite{zhou2020transforming} and StarGAN~\cite{rizos2020stargan}. We reproduce the CopyPaste~\cite{pappagari2021copypaste} method by randomly concatenating two emotional utterances with the same emotion, which corresponds to the Same Emotion CopyPaste (SE-CP) setting in \cite{pappagari2021copypaste}. Additionally, we perturb speech on speed with the factors of 0.9, 1.0 and 1.1. Pitch shift is adopted by randomly raising or lowering 2 semitones on each audio. Table~\ref{table:comparison-with-other} shows that EmoAug outperforms previous methods by a big margin.



\vspace{-0.2cm}
\begin{table}[h]
\centering
\caption{Results of SER on IEMOCAP dataset with 5-fold cross-validation and leave-one-session-out settings.}
\resizebox{7cm}{!}{
\begin{tabular}{lcc}
\toprule
\textbf{Methods} & \textbf{WA} & \textbf{UA}\\\hline
\multicolumn{1}{l}{\textbf{Supervised Methods}}&&\\\hline
CNN-ELM+STC attention~\cite{guo2021representation}& 61.32 &60.43\\

IS09-classification~\cite{tarantino2019self} & 68.10 & 63.80\\
Co-attention-based fusion~\cite{zou2022speech} & 69.80 & 71.05 \\\hline
\multicolumn{1}{l}{\textbf{Self-supervised Methods}}&&\\\hline

Data2Vec Large~\cite{baevski2022data2vec} & 66.31&- \\
WavLM Large~\cite{chen2022wavlm} & 70.62&- \\
HuBERT Large & 70.24& 71.13 \\\hline
\multicolumn{1}{l}{\textbf{Emotional Voice Conversion Methods}} &     \\\hline
HuBERT Large $+$ CycleGAN~\cite{zhou2020transforming}& 71.57 & 72.02 \\
HuBERT Large $+$ StarGAN~\cite{rizos2020stargan}& 71.51 & 72.13 \\\hline

\multicolumn{1}{l}{\textbf{Data Augmentation Methods}} &     \\\hline
VTLP~\cite{etienne2018cnn} & 66.90 & 65.30\\
HuBERT Large $+$ CopyPaste & 70.79& 71.35 \\
HuBERT Large $+$ Speed Perturbation & 70.35& 71.19 \\
HuBERT Large $+$ Pitch Shift & 70.47& 71.24 \\\hline

\multicolumn{1}{l}{\textbf{Our Methods}} &     \\\hline
HuBERT Large $+$ EmoAug& \textbf{72.66} &\textbf{73.75}\\\bottomrule
\end{tabular}}
\label{table:comparison-with-other}
\vspace{-0.5em}
\end{table}

\vspace{-0.4cm}
\section{Conclusion}
\vspace{-0.2cm}





We introduce EmoAug to tackle the challenges of data scarcity and data imbalance in SER. EmoAug is composed of a semantic encoder, a paralinguistic encoder, and a decoder to reconstruct speech in an unsupervised manner. Speaking style transfer is performed by altering the input of the paralinguistic encoder.
Experimental results on the
IEMOCAP dataset suggest that EmoAug can effectively enrich
emotion expressions with different stress patterns, rhythms
and intensities, and achieve superior performance in SER compared to
state-of-the-art supervised and self-supervised methods.


\vspace{-0.3cm}
\section{Acknowledgement}
\vspace{-0.1cm}
This work was supported in part by the National Science and Technology Major Project of China (2021ZD0114303), in part by the Youth Foundation Project of Zhejiang Lab (K2023KH0AA01), in part by the CML Project funded by the DFG, and in part by the Philosophy and Social Science Training Project of Xinjiang University (23CPY049).

\small
\newcommand{\BIBdecl}{\setlength{\itemsep}{0.5cm}}
\bibliographystyle{IEEEbib}
\bibliography{main}

\end{document}